\documentclass[10pt,reqno,twoside,paper=letter]{amsart}

\usepackage[letterpaper,left=20mm,right=14mm,layoutvoffset=10mm]{geometry}
\usepackage{epsfig,graphicx}
\usepackage{epstopdf}

\usepackage[bf,small]{titlesec}
\titlelabel{\thetitle.\,\,\,}
\usepackage{amssymb,amsfonts}
\usepackage{amsmath}
\usepackage{float} 
\usepackage{multicol}
\usepackage{fancyhdr}
\usepackage[latin1]{inputenc}				
\usepackage[comma,longnamesfirst,sectionbib]{natbib}
\usepackage{enumerate}
\usepackage[symbol*]{footmisc}

\newcommand{\inforevista}{\scriptsize  Rev. Acad. Colomb. Cienc. Ex. Fis. Nat. nn(nnn):ww--zzz,ddd-ddd de 2016}
\begin{document}
%
%
\pagenumbering{arabic}
\fancypagestyle{plain}{%
\fancyhf{} 
\fancyfoot[R]{\thepage} %
\fancyhead[L]{\inforevista}
\renewcommand{\headrulewidth}{0pt}
\renewcommand{\footrulewidth}{0pt}}
\thispagestyle{plain} 
\pagestyle{fancy}     
\fancyhead{} 
\renewcommand{\headrulewidth}{0.0pt} 
\fancyhead[LO]{\inforevista}
\fancyhead[RO]{\scriptsize Stability of Axisymmetric Relativistic Thin Disks}
\fancyhead[LE]{\scriptsize E. A. Becerra, F. L. Dubeibe, G. A. González}
\fancyhead[RE]{\inforevista}
\fancyfoot{} 
\fancyfoot[LE]{\thepage}
\fancyfoot[RO]{\thepage}

\begin{flushright}
\rule[-0.5ex]{0.5ex}{3.0ex} {\large Physical Sciences}
\end{flushright}
\vspace*{0.3cm}

\begin{center}
{\LARGE \textbf{On the Influence of the Mass Definition in the Stability of Axisymmetric Relativistic Thin Disks \\}}
\end{center}

\vspace{3mm}
\begin{center}
\textbf{\small Eduar A. Becerra${}^{1,}$, Fredy L. Dubeibe${}^{1,2}$, Guillermo A. González${}^{1,}\footnotemark[1]$}
\vspace{3mm}

{\scriptsize
${}^{1}$Escuela de Física, Universidad Industrial de Santander, Bucaramanga, Colombia\\
${}^{2}$Facultad de Ciencias Humanas y de la Educación, Universidad de los Llanos, Villavicencio, Colombia\\
 }
\end{center}
\footnotetext[1]{Correspondence: G. A. González, guillermo.gonzalez@saber.uis.edu.co, Received xxxxx XXXX; Accepted xxxxx XXXX.}

\begin{Small}
\vspace{3.0mm}
\rule{\textwidth}{0.4pt}

\begin{center}
\begin{minipage}{14cm}
\vspace{3mm}

\textbf{Abstract}
\vspace{3mm}

{The study on the stability of relativistic disks is one of the most important criteria for the characterization of astrophysically relevant galactic or accretion disks models. In this paper, we perform an analysis of the stability of static axisymmetric relativistic thin disks, by introducing a first-order perturbation into the energy-momentum tensor of the fluid. The formalism is applied to three particular models built with the aid of the displace-cut-reflect (DCR) method, and previously considered in literature \citep{Ujevic2004}, but modifying the mass criteria, {\it i.e.}, using the Komar mass instead of the total surface mass. Under this conditions, it is found that the total mass values are independent of the parameters of the DCR-method, which let us choose the boundary condition for the cutoff radius, such that it takes the maximum value that allows an appreciable and well-behaved perturbation on the disk. As a general result, we found that the Komar mass is more appropriate to define the cutoff radius.
\\[1mm]

\textbf{Key words:}  General Relativity, Relativistic thin disks, Stability.}

\vspace{3mm}

\textbf{Influencia de la definici\'on de masa en la estabilidad de discos relativistas axialsimétricos }
\vspace{3mm}

\textbf{Resumen}
\vspace{3mm}

{Uno de los criterios más importantes para la caracterización de modelos galácticos o discos de acreción astrofísicamente relevantes, es el análisis de la estabilidad de dichos modelos. En este trabajo, se realiza un análisis de la estabilidad de discos delgados estáticos relativistas con simetría axial, mediante la introducción de una perturbación de primer orden en el tensor de energía-impulso del fluido. El formalismo se aplica a tres modelos construidos con el método de desplazamiento-corte-reflexión (DCR), previamente considerados en la literatura \citep{Ujevic2004}, pero modificando el criterio de masa, es decir, usando la masa de Komar en lugar de la masa total superficial. Bajo estas condiciones, se encuentra que los valores de masa total son independientes de los parámetros del método DCR, lo que permite elegir la condición de frontera para el radio de corte que tome el valor máximo y a la vez permita una perturbación apreciable y bien comportada en el disco. Como resultado general, se encuentra que para la mayor\'ia de modos de oscilaci\'on, la masa de Komar es mas apropiada para definir el radio de corte.
\\[1mm]

\textbf{Palabras clave:} Relatividad  General, Discos delgados Relativistas, Estabilidad.}
\end{minipage}
\end{center}

\rule{\textwidth}{0.4pt}
\end{Small}

\begin{small}
\columnsep 0.5 cm
\begin{multicols}{2}

\setlength{\parskip}{.3cm}

\section*{Introduction}

During the last decades, considerable efforts have been made to obtain exact analytical solutions suitable to modeling axisymmetric thin disks, within the framework of Newton's and Einstein's theories of gravity. Such models are of astrophysical interest because they can be used to  model accretion disks, galaxies in thermodynamic equilibrium or galaxies with black holes centers \citep{ bicak1993-2, ledvinka1998}. Moreover, the addition of electromagnetic fields in those space-times allows studying neutron stars formation, white dwarfs, and quasars \citep{munoz2011, alpar2001}. 

Since the seminal works on exact solutions representing static thin disks carried out by Bonnor and Sackfield \citep{bonnor1968} and Morgan and Morgan \citep{morgan1969}, more realistic models have been proposed \citep{pichon1996, gonzalez2000}. The superposition of static and stationary thin disks with black holes at the center, has been considered by Lemos, Letelier and Semerak \citep{lemos1993, semerak2002, semerak2004}. Vogt and Letelier studied the inclusion of electromagnetic fields into thin disks made of dust \citep{vogt2004} and of charged perfect fluid \citep{vogt2004-2}. Also interesting is the case of thick disks proposed by Gonz\'alez and Letelier \citep{gonzalez2004}, who extended the DCR method to include thick disks in their models.\footnote{For the interested reader, a complete review of the state-of-the-art on relativistic disks was made by \cite{Semerak-Book-2002}.} 

Stability is an essential criterion to determine whether or not a model can be applied to describe an astrophysical system present in nature. In general, there are two approaches to study the stability of relativistic disks: The first option is based on analyzing the stability of particle orbits along geodesics (see {\it e.g.,} \citep{letelier2003} and \citep{vogt2003}), while the second option consists in perturbing the energy-momentum tensor (see {\it e.g.,} \citep{seguin1975}). From a theoretical point of view, the latter option is more rigorous, because in this case the collective behavior of the particles is taken into account. Working on this line, \cite{Ujevic2004} investigated the stability of three particular models for relativistic thin disks, performing a first-order perturbation analysis with variable coefficients. However, as a pathological result, the authors find that the total mass of the disk depends on the parameters of the DCR method, such that the boundary conditions are also dependent on these parameters. 

With the aim to avoid the undesired dependencies between parameters and to observe the possible changes in the stability, in the present paper we redo the calculations made by \cite{Ujevic2004} using the Komar mass definition instead of the mass definition (along the paper we will call it total surface mass) introduced in \cite{vogt2003}. The new results let us to choose the boundary condition for the cutoff radius such that it takes the maximum value allowing an appreciable and well-behaved perturbation on the disk.

\section*{Derivation of First-Order Perturbation Equations}\label{dischalo}

Following \cite{Ujevic2004}, and for the sake of self-consistency of the current paper, in what follows we present the derivation of the first-order perturbation equations for relativistic thin disks. Let us start considering that the energy-momentum tensor for an isotropic fluid with a discoid shape and without heat flow can be written as
\begin{equation}
\label{19}
T^{\mu \nu} = Q^{\mu \nu} \delta\left(z\right),
\end{equation}
where $\delta$ denotes the Dirac delta function and 
\begin{equation*}
Q^{\mu \nu} = \sigma U^{\mu}U^{\nu} + p_{r}X^{\mu}X^{\nu} + p_{\varphi}Y^{\mu}Y^{\nu},
\end{equation*}
with $\sigma$ the surface energy density, $p_{r}$ and $p_{\varphi}$ the radial and azimuthal pressure, respectively, and $U^{\mu}, X^{\mu}$, and $Y^{\mu}$ the non-zero components of the orthonormal tetrad. Assuming that the first-order perturbations in the Einstein field equations do not modify the background metric, the perturbed equation for the energy-momentum tensor reads as 
\begin{equation}
\label{12}
\left(\delta T^{\mu \nu}\right)_{; \mu} = 0.
\end{equation}
Introducing the definition of energy-momentum tensor for a thin disk \eqref{19}, in the perturbed equation \eqref{12}, and integrating with respect to the coordinate $z$, we find 
 \begin{equation}
\label{22}
\int \left\lbrace \left( \delta Q^{\mu \nu} \right)_{;\mu} \delta\left(z\right) + \delta Q^{z \nu}\left[\delta\left(z\right)\right]_{,z} \right\rbrace \sqrt{g_{zz}} dz= 0.
\end{equation}
As a consecuence of the DCR method, the metric components should only depend on $r$ and $|z|$ \citep{vogt2003}. Additionally, if we define the value of $|z|_{,z}$ at $z = 0$ equals zero, the perturbed equation reduces to
\begin{equation}
\label{25}
\left( \delta Q^{\mu \nu} \right)_{;\mu}\biggl| _{z=0} = 0.
\end{equation}
Taking into account that the perturbed vectors must satisfy the orthonormality condition, and assuming that $\delta X^{\varphi}=0$ (since the four-velocity and the thermodynamical variables do not depend on this quantity), we obtain 
\begin{equation}
\begin{aligned}\label{deltas}
& \delta Y^{r} = \delta U^{t} = \delta X^{r} = \delta Y^{\varphi} = 0,\\ 
& \delta X^{t} = - \dfrac{X_{r}}{U_{t}}\delta U^{r}, \ \delta Y^{t} = - \dfrac{Y_{\varphi}}{U_{t}}\delta U^{\varphi}.
\end{aligned}
\end{equation}
Due to the fact that the metric is static and axisymmetric, and the lack of $z$-dependences in $T^{\mu \nu}$, all the coefficients depend only on the radial coordinate; therefore, the general perturbation can be chosen as 
\begin{equation}
\label{28}
\delta \xi^{\mu}\left(t, r, \varphi\right) = \delta \xi^{\mu}\left(r\right) e^{i\left(k \varphi - wt\right)},
\end{equation}
with $k$ the wave number and $w$ the angular frequency of the perturbation. We will focus on the term $\delta\xi^{\mu}\left(r\right)$, which has a significant role for the system stability,. 

Therefore, after substituting the perturbation \eqref{28} into equation \eqref{25}, and replacing the conditions \eqref{deltas} in the resulting expression, the respective perturbed equations for $t$, $r$, and $\varphi$, read as
\begin{equation}
\label{29}
\begin{small}
\begin{aligned}
& \delta U^{r}_{\ ,r}\left(\sigma U^{t} - \dfrac{p_{r}}{U_{t}}\right) + \delta U^{r}\left\lbrace \left(\sigma U^{t}\right)_{,r} +  \right.\\
& \left. \sigma U^{t} \left(2\Gamma^{t}_{\ tr} + \Gamma ^{\mu}_{\ \mu r}\right) - \left(\dfrac{X_{r}}{U_{t}}\right)_{,r} p_{r}X^{r} -\right.\\\
& \left. \dfrac{X_{r}}{U_{t}}\left[\left(p_{r} X^{r}\right)_{,r} + p_{r} X^{r}\left(2\Gamma^{t}_{\ tr} + \Gamma ^{\mu}_{\ \mu r} \right)\right]\right\rbrace + \\
&  \delta U^{\varphi} \left[ik\left(\sigma U^{t} - \dfrac{p_{\varphi}}{U_{t}}\right)\right] - \delta \sigma \left(iw U^{t}U^{t}\right)= 0,
\end{aligned}
\end{small}
\end{equation}

\begin{equation}
\label{30}
\begin{aligned}
&  \delta U^{r}\left[iw\left(\dfrac{p_{r}}{U_{t}} - \sigma U^{t}\right) \right] + \delta \sigma \left(U^{t}U^{t}\Gamma ^{r}_{\ tt}\right) + \\
& \delta p_{r}\left[\left(X^{r} X^{r}\right)_{,r} + X^{r} X^{r}\left(\Gamma^{r}_{\ rr} + \Gamma ^{\mu}_{\ \mu r} \right)\right] +\\
& \delta p_{r,r}\left(X^{r} X^{r}\right) + \delta p_{\varphi} \left(Y^{\varphi}Y^{\varphi}\Gamma ^{r}_{\ \varphi \varphi}\right) = 0,
\end{aligned}
\end{equation}

\begin{equation}
\label{31}
\begin{aligned}
&\delta U^{\varphi}\left[w\left(\dfrac{p_{\varphi}}{U_{t}} - \sigma U^{t}\right) \right] + \delta p_{\varphi} \left(k Y^{\varphi}Y^{\varphi}\right) = 0,
\end{aligned}
\end{equation}

Finally, from the set of differential equations presented above, and the equation of state of a perfect fluid, $\delta p= \delta \sigma (p_{,r}/\sigma_{,r})$, the differential equation for the perturbed energy density takes the form
\begin{equation}
\label{68}
A\delta \sigma_{,rr}+B\delta \sigma_{,r}+C\delta \sigma=0,
\end{equation}
whose coefficient $A$, $B$, and $C$, are given by
\begin{equation*}
A=-\dfrac{A_{1} B_{1}}{B_{2}}\left(\dfrac{p_{,r}}{\sigma_{,r}}\right),
\end{equation*}
\begin{equation*}
\begin{aligned}
B=& A_{1}\left\lbrace \left(\dfrac{p_{,r}}{\sigma_{,r}}\right)\left[\dfrac{B_{1} B_{2,r}}{B^{2}_{2}}-\dfrac{B_{1,r}+B_{4}+B_{5}}{B_{2}}\right]-\dfrac{B_{3}}{B_{2}}- \right. \\
& \left. \dfrac{2 B_{1}}{B_{2}}\left(\dfrac{p_{,r}}{\sigma_{,r}}\right)_{,r}\right\rbrace-\dfrac{A_{2} B_{1}}{B_{2}}\left(\dfrac{p_{,r}}{\sigma_{,r}}\right),
\end{aligned}
\end{equation*}
\begin{equation*}
\begin{aligned}
C=& A_{1}\left\lbrace \left(\dfrac{p_{,r}}{\sigma_{,r}}\right)_{,r}\left[\dfrac{B_{1} B_{2,r}}{B^{2}_{2}}-\dfrac{B_{1,r}+B_{4}+B_{5}}{B_{2}}\right]-\dfrac{B_{3,r}}{B_{2}}+\right.\\
& \left. \left(\dfrac{p_{,r}}{\sigma_{,r}}\right)\left[\dfrac{B_{2,r}}{B^{2}_{2}} \left(B_{4}+B_{5}\right)-\dfrac{B_{4,r}+B_{5,r}}{B_{2}}\right]+\right.\\
& \left. \dfrac{B_{2,r} B_{3}}{B^{2}_{2}}-\dfrac{B_{1}}{B_{2}}\left(\dfrac{p_{,r}}{\sigma_{,r}}\right)_{,rr}\right\rbrace -\dfrac{A_{3} C_{2}}{C_{1}}\left(\dfrac{p_{,r}}{\sigma_{,r}}\right)-A_{4}\\
&  -\dfrac{A_{2}}{B_{2}}\left\lbrace B_{1}\left(\dfrac{p_{,r}}{\sigma_{,r}}\right)_{,r}+\left(\dfrac{p_{,r}}{\sigma_{,r}}\right) \left[B_{4}+B_{5}\right]+B_{3}\right\rbrace,
\end{aligned}
\end{equation*}
where $A_{i}$,  $B_{i}$, and  $C_{i}$, are the factors multiplying the perturbed variables in equations \eqref{29},  \eqref{30}, and \eqref{31}, respectively, with the indexes in numerical order according to their order of appearance.

Due to the cumbersome form of equation \eqref{68}, it must be solved numerically. For this purpose, we shall impose Dirichlet boundary conditions, one at the center of the disk and the other one at the final boundary of the domain. However, it should be noted that the disk has an infinite radial extension, for this reason, it is necessary to introduce a cutoff on the radial coordinate. 

\section*{Thermodynamic variables}\label{partmodel}

In the previous section, we explicitly wrote down the perturbed equations for thin disks, nevertheless, such expressions are given in terms of the thermodynamic variables of the fluid, which requires finding explicit formulas for the surface energy density and the pressures on the disk. To this end, let us represent the matter distribution on a hypersurface $\Sigma$, defined by the function $l(x^{\alpha})=z$, which divides the space-time into two regions: $M^{+}$ on top and $M^{-}$ at the bottom. Therefore, the normal vector to the hypersurface $\Sigma$ is given by $n_{\alpha}=l_{,\alpha}=\delta^{z}_{\alpha}$, and the components of the metric tensor must satisfy
\begin{equation}
\label{32}
g^{-}_{\mu \nu} (r, z)= g^{+}_{\mu \nu} (r, -z),
\end{equation} 
such that
\begin{equation}
\label{33}
g^{-}_{\mu \nu, z} (r, z)= -g^{+}_{\mu \nu, z} (r, -z),
\end{equation}
where $g^{+}_{\alpha \beta}$ and $g^{-}_{\alpha \beta}$, should be understood as the metric tensors for the regions defined by  $z>0 \ (M^{+})$ and $z<0 \ (M^{-})$, respectively. 

By taking the limit $z\rightarrow 0$, the discontinuities in the first derivatives of the metric tensor take the form
\begin{equation}
\label{35}
b_{\mu \nu}=[g_{\mu \nu, z}]=g^{+}_{\mu \nu,z}\biggl| _{z=0} - \ \ g^{-}_{\mu \nu,z}\biggl| _{z=0} = 2 g^{+}_{\mu \nu, z} \biggl| _{z=0} ,
\end{equation}
with $b_{\mu \nu}$, the jump of the first derivative through $\Sigma $. Using the distributions method  \citep{papapetrou1968,toolkit}, the metric can be written as 
\begin{equation}
\label{36}
g_{\alpha \beta} =  \Theta\left(l\right)g^{+}_{\alpha \beta} + \Theta\left(-l\right)g^{-}_{\alpha \beta},
\end{equation}
where $\Theta\left(l\right)$ is the usual Heaviside function. Performing the derivative of equation \eqref{36} we obtain
\begin{equation}
\label{37}
g_{\alpha \beta, \gamma} =  \Theta\left(l\right)g^{+}_{\alpha \beta, \gamma} + \Theta\left(-l\right)g^{-}_{\alpha \beta, \gamma} + n_{\gamma}\delta (l) \left[ g_{\alpha \beta}\right].
\end{equation}

The last term in the right-hand side of equation \eqref{37} is singular, consequently the Christoffel symbols would not be defined as a distribution. An alternative way to avoid this difficulty, is to impose that the metric is continuous on the hypersurface, {\it i.e.}, $\left[ g_{\alpha \beta}\right] =  g^{+}_{\alpha \beta} - g^{-}_{\alpha \beta} = 0$, such that
\begin{equation}
\label{39}
\Gamma ^{\alpha}_{\ \beta \gamma} =   \Theta\left(l\right)\Gamma ^{\alpha +}_{\ \beta \gamma} + \Theta\left(-l\right)\Gamma ^{\alpha -}_{\ \beta \gamma},
\end{equation}
whose derivative reads as
\begin{equation}
\label{40}
\Gamma ^{\alpha}_{\ \beta \gamma , \delta} =   \Theta\left(l\right)\Gamma ^{\alpha +}_{\ \beta \gamma , \delta} + \Theta\left(-l\right)\Gamma ^{\alpha -}_{\ \beta \gamma , \delta} + n_\delta (l)[\Gamma ^{\alpha}_{\ \beta \gamma}].
\end{equation}
The corresponding Riemann curvature tensor is given by
\begin{equation}
\label{42}
R ^{\alpha}_{\ \beta \gamma \delta} =   \Theta\left(l\right)R^{\alpha +}_{\ \beta \gamma \delta} + \Theta\left(-l\right)R ^{\alpha -}_{\ \beta \gamma \delta} + \delta(l) \widehat{R}^{\alpha }_{\ \beta \gamma \delta},
\end{equation}
where $R^{\alpha \pm}_{\ \beta \gamma \delta}$  are the tensors defined in $M^{\pm}$ and
\begin{equation}
\label{43}
\widehat{R}^{\alpha }_{\ \beta \gamma \delta} = [\Gamma ^{\alpha}_{\ \beta \delta}]n_{\gamma}-[\Gamma ^{\alpha}_{\ \beta \gamma}]n_{\delta},
\end{equation}
with $\ [\Gamma ^{\alpha}_{\ \beta \gamma}]=\dfrac{1}{2}\left(b^{\alpha}_{\ \beta} n_{\gamma}+b^{\alpha}_{\ \alpha} n_{\beta}-b_{\beta \gamma} n^{\alpha}\right)$.

From the last expression, it is clear that Riemann tensor, the Ricci tensor, and Ricci scalar on the hypersurface are 
\begin{equation}
\label{46}
\begin{aligned}
&\widehat{R}^{\alpha }_{\ \beta \gamma \delta}=\dfrac{1}{2}\left\lbrace b^{\alpha}_{\ \delta}n_{\beta} n_{\gamma}-b^{\alpha}_{\ \gamma}n_{\beta} n_{\delta}+b_{\beta \gamma}n^{\alpha} n_{\delta}-b_{\beta \delta}n^{\alpha} n_{\gamma}\right\rbrace,\\
&\widehat{R}_{\beta \delta}=\dfrac{1}{2}\left\lbrace b^{\alpha}_{\ \delta}n_{\beta} n^{\alpha}-b^{\alpha}_{\ \alpha}n_{\beta} n_{\delta}+b_{\beta \alpha}n^{\alpha} n_{\delta}-b_{\beta \delta}n^{\alpha} n_{\alpha}\right\rbrace,\\
& \widehat{R}= \left\lbrace b_{\mu \alpha}n^{\alpha} n^{\mu}-b^{\alpha}_{\ \alpha}n^{\alpha} n_{\alpha}\right\rbrace.
\end{aligned}
\end{equation}
On the other hand, the energy-momentum tensor $T_{\beta \delta}$ can be expressed as
\begin{equation}
\label{47}
T_{\beta \delta}= \Theta\left(l\right)T^{+}_{\beta \delta} + \Theta\left(-l\right)T ^{-}_{\beta \delta} + \delta(l) Q_{\beta \delta},
\end{equation}
where $Q_{\beta \delta}$ is the energy-momentum tensor associated with the hypersurface, and $T ^{\pm}_{\beta \delta}$ are the energy-momentum tensors associated to $M^{\pm}$. Hence, we can write the Einstein field equations as follows
\begin{equation}
\label{48}
\begin{aligned}
&R^{\pm }_{\beta \delta}-\dfrac{1}{2}g_{\beta \delta}R^{\pm} =\kappa T ^{\pm}_{\beta \delta}\\
&\widehat{R}_{\beta \delta}-\dfrac{1}{2}g_{\beta \delta}\widehat{R}=\kappa Q_{\beta \delta}.
\end{aligned}
\end{equation} 
Given that the only source of gravitational field is a thin distribution of matter, the energy-momentum tensor satisfies 
\begin{equation}
\label{49}
\begin{aligned}
\kappa Q_{\beta \delta}=&\dfrac{1}{2}\left\lbrace b_{\alpha \delta}n_{\beta} n^{\alpha}-b^{\alpha}_{\ \alpha}n_{\beta} n_{\delta}+b_{\beta \alpha}n^{\alpha} n_{\delta}- b_{\beta \delta}n^{\alpha} n_{\alpha} \right.\\
& \left. - \left( b_{\mu \alpha}n^{\alpha} n^{\mu}-b^{\alpha}_{\ \alpha}n^{\alpha} n_{\alpha}\right)g_{\beta \delta}\right\rbrace.
\end{aligned}
\end{equation}
which, in quasi-cylindrical coordinates $x^{\alpha}=(t, r, \varphi, z)$ and considering the hypersurface $\Sigma$ defined by the function $l(x^{\alpha})=z$, with normal vector $n_{\alpha}=\delta^{z}_{\alpha}$, takes the form
\begin{equation}
\label{50}
\begin{aligned}
\kappa Q^{\mu}_{\ \beta}=& \dfrac{1}{2}\left\lbrace b^{z \mu}\delta^{z}_{\beta}-g^{\mu z}\delta^{z}_{\beta}b^{\alpha}_{\ \alpha}+g^{\mu z} b^{z}_{\ \beta}- g^{zz}b^{\mu}_{\ \beta}  \right.\\
& \left. - \left( b^{zz} -g^{zz}b^{\alpha}_{\ \alpha}\right)\delta^{\mu}_{\beta}\right\rbrace.
\end{aligned}
\end{equation}
From the above equation, and as noted by \cite{vogt2003}, it can be shown that the non-zero components of the surface energy-momentum tensor are
\begin{equation}
\label{51}
Q^{t}_{\ t}=\sigma=\dfrac{1}{16 \pi} g^{zz} \left( b^{r}_{\ r} + b^{\varphi}_{\ \varphi}\right),
\end{equation}
\begin{equation}
\label{52}
Q^{r}_{\ r}=p_{r}=\dfrac{1}{16 \pi} g^{zz} \left( b^{t}_{\ t} + b^{\varphi}_{\ \varphi} \right),
\end{equation}
\begin{equation}
\label{53}
 Q^{\varphi}_{\ \varphi}=p_{\varphi}=\dfrac{1}{16 \pi} g^{zz} \left( b^{t}_{\ t} + b^{r}_{\ r} \right).
\end{equation}

\section*{Mass Definition}

The mass concept in general relativity is not unique and there are several different definitions that are applicable under different circumstances. In the case of stationary spacetimes, the commonly accepted definitions are: the total volumetric (or surface) mass, Komar, ADM and Bondi-Sach masses. However, it is a well-known fact that for stationary spacetimes the ADM and Bondi-Sach masses are exactly alike \citep{toolkit}. Besides, in the asymptotically flat case, it can be shown that the ADM and the Komar masses are equivalent \citep{relativity}. Then, in the stationary, axisymmetric, asymptotically flat, vacuum solutions of Einstein's equation, there are only two possibilities to choose from, the total volumetric (or surface) mass or the Komar mass. As commented in the introduction, the main difference between the present paper and the one by \cite{Ujevic2004}, is that they used the total surface mass, while along this paper, we use the Komar mass definition, which is independent of the methods used to build the particular solutions of Einstein's equation.

A necessary condition for the definition of mass, is that must not involve any dependence with the specific choice of coordinates. This property is achieved for stationary and axially symmetric spacetimes, through the Komar formula $M_{k}$, which reads as
\begin{equation}
\label{74}
M_{k}=2 \int_\Sigma \left(T_{\alpha \beta}-\dfrac{1}{2}Tg_{\alpha \beta} \right)n^{\alpha}\xi^{\beta}_{(t)} \sqrt{h} \ d^{3}y,
\end{equation} 
where $\Sigma$ is spacelike hypersurface, $n^{\alpha}$ is timelike  vector normal to $\Sigma$, \ $\xi^{\beta}_{(t)}$ is timelike Killing vector and $h$ is the determinant of metric associated to $\Sigma$. For a static and diagonal metric, the following relations holds
\begin{equation*}
n^{\alpha}=-\dfrac{g^{\alpha \beta}}{\sqrt{|g^{tt}|}} \ \delta^{t}_{\beta},  \quad \quad   \xi^{\beta}_{(t)}=\delta^{\beta}_{t}, \quad \quad  h=g_{zz} \ g_{rr} \ g_{\varphi \varphi},
\end{equation*}
such that the Komar mass takes the form
\begin{equation}
\label{75}
M_{k}=\int_0^{2\pi} \int_0^\infty \left(\sigma+p_{r}+p_{\varphi}\right)\sqrt{\lvert g_{tt}\lvert} \sqrt{g_{zz}}\sqrt{g_{rr}} \sqrt{g_{\varphi \varphi}} \ dr d\varphi.
\end{equation}

The Komar mass definition differs from the total surface mass in the fact that the former one considers all the contributions to the energy-momentum tensor, while the surface mass, given by
\begin{equation}
\label{bar}
M_{s}=\int_0^{2\pi} \int_0^\infty \sigma \sqrt{g_{zz}}\sqrt{g_{rr}} \sqrt{g_{\varphi \varphi}} \ dr d\varphi,
\end{equation}
takes only into account the surface energy density.

\section*{Case 1: Isotropic Schwarzschild Thin Disk}\label{fitting}

The isotropic Schwarzschild thin disk metric in quasi-cylindrical Weyl-Papapetrou coordinates was obtained by \cite{vogt2005}, and can be written as
\begin{equation}
\label{56}
ds^{2}=-\left(\dfrac{2R-m}{2R+m}\right)^{2} dt^{2}+ \left(1+ \dfrac{m}{2R}\right)^{4}(dr^{2}+ r^{2}d\varphi^{2}+dz^{2})
\end{equation}
where $m$ is a positive constant and $R^{2}= r^{2}+\left(\lvert z \rvert+ a\right)^{2}$. 

Thus, the corresponding expressions for the pressure components and surface energy density (see {\it e.g.,} \cite{vogt2003}), are obtained by using equations (\ref{51}-\ref{53}),  
\begin{equation}
\label{58}
\sigma= \dfrac{16 m a R^{2}_{0}}{\pi (2R_{0} + m)^{5}},
\end{equation}
\begin{equation}
\label{59}
p= \dfrac{8 m^{2} a R^{2}_{0}}{\pi (2R_{0} + m)^{5}(2R_{0} - m)},
\end{equation}
with $R_{0}=R(r,z=0)$ and $p=p_{r}=p_{\varphi}$. 

Let us define a general cutoff radius as the radial distance at which the matter within the thin disk formed up to such radius corresponds to the $n\%$ of the total matter of the infinite disk, {\it i.e.}
\begin{equation}
n M_{T}=M_{r_{c}},
\end{equation}
with $M_{T}$ the total mass of the infinite disk and $M_{r_{c}}$ the mass up to the cutoff radius. As noted in the previous section, the total mass value may depend on the considered definition, such that the cutoff radius could depend also on this choice. From \eqref{bar}, (\ref{56}) and (\ref{58}), and in accordance with Eq. (12) of \cite{vogt2003}, we find
\begin{equation}
\label{Mb}
M_{s}=m\left(1+\frac{m}{4a}\right),
\end{equation}
and
\begin{equation}
\label{Mrc}
M_{r_{c}}=m\left(1+\frac{m}{4a}\right)-\frac{m a}{2 r_{c}}\left(2+\frac{m}{2 r_{c}}\right).
\end{equation}
On the other hand, from the Komar mass definition \eqref{75} and using (\ref{56}), (\ref{58}) and (\ref{59}), we get
\begin{equation}
\label{Mk}
M_{k}=m,
\end{equation}
and
\begin{equation}
\label{Mrc}
M_{r_{c}}=m-\frac{m a}{r_{c}}.
\end{equation}
Hence, the resulting expression for the cutoff radius when considering Eq. \eqref{bar} is
\begin{equation}\label{eq:rcUL}
r_{c}^{*}=\frac{2 a^2+a\sqrt{4 a^2 + m(1-n)(4a+m)}}{(1-n)(4a+m)},
\end{equation}
while using Eq. \eqref{75}, we get
\begin{equation}\label{eq:rcN}
r_{c}=\frac{a}{1-n}
\end{equation} 
In Fig. 1 we show a comparison between the cutoff radii $r_{c}$ and $r_{c}^{*}$, using different values of the parameters $m$ and $a$. Taking into account that values of $n$ smaller than $0.9$ are not used nor physically appropriate, we plot the difference  $r_{c}-r_{c}^{*}$ in the range $n\in[0.9,1]$.
\begin{figure}[H]
\begin{center}
\includegraphics[scale=0.85]{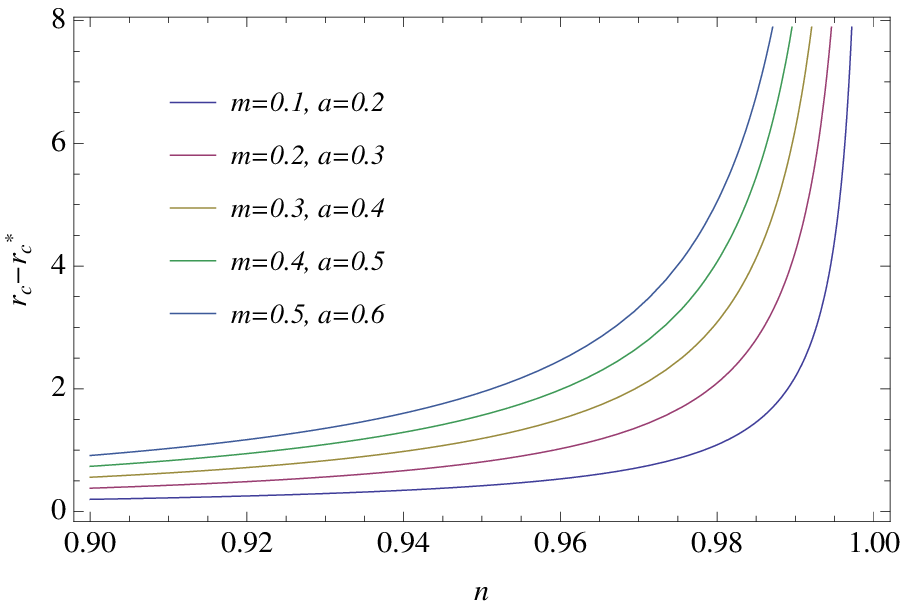}
\label{fig1n}
\end{center}
\end{figure}
\vspace{-8mm}
\begin{small}
FIGURE 1. Comparison between the cutoff radii $r_{c}$ and $r_{c}^{*}$ using different values of the parameters $m$ and $a$.
\end{small}

In all that follows, except where especially noted, we define the cutoff radius $r_{c}$ using the 95\% of the total mass of infinite disk,\footnote{This percentage corresponds to the optimal value to get a non-negligible perturbation on the disk.}  and the remaining 5\% of matter is distributed along the plane $z=0$ from $r_{c}$ to infinity. Moreover, we assume that at $r=0$, the perturbation equals 10\% of the unperturbed energy density value and at $r=r_{c}$ the perturbation vanishes, {\it i.e.} $\delta \sigma(r=r_{c})=0$. 

By setting $m=0.5$ and $a=0.6$, which are the set of parameters that best fit the expected energy density profile in a realistic model, we find $r_{c}\approx 12$ and $r_{c}^{*}\approx 10$. In Fig. 2 we present the numerical solution to the differential equation \eqref{68} using the Komar mass definition (upper panel) and the total surface mass (lower panel). In this figure, we show the perturbed energy density profile for different values of $w$. It can be seen that the parameter $w$ is proportional to the frequency of oscillations within the disk, {\it i.e.} the number of ring-like structures increases with increasing $w$. Also, it is worth noting that the oscillations quickly decay to zero regardless of the oscillation modes, however the oscillation amplitude is smaller for Eq. \eqref{75} (upper panel) than for Eq. \eqref{bar}  (upper panel). 
\begin{figure}[H]
\begin{center}
\includegraphics[scale=0.85]{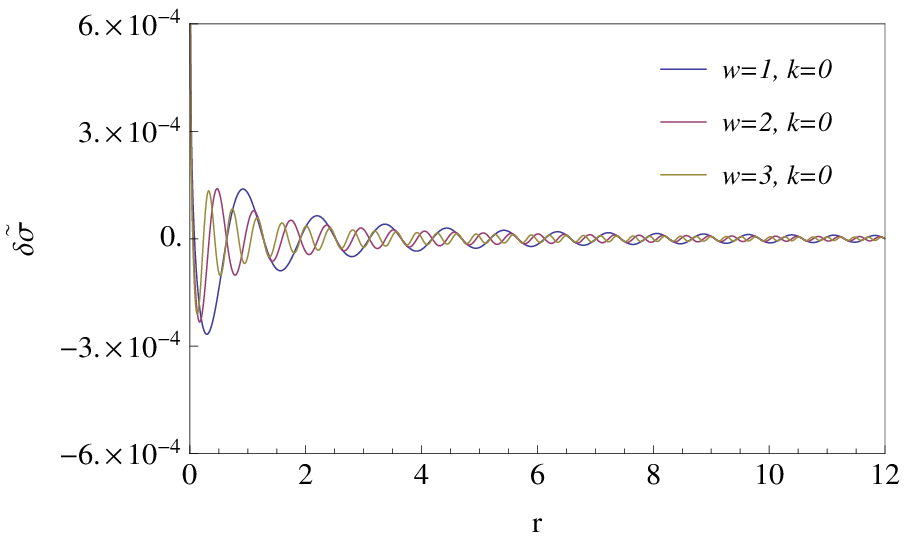}
\includegraphics[scale=0.85]{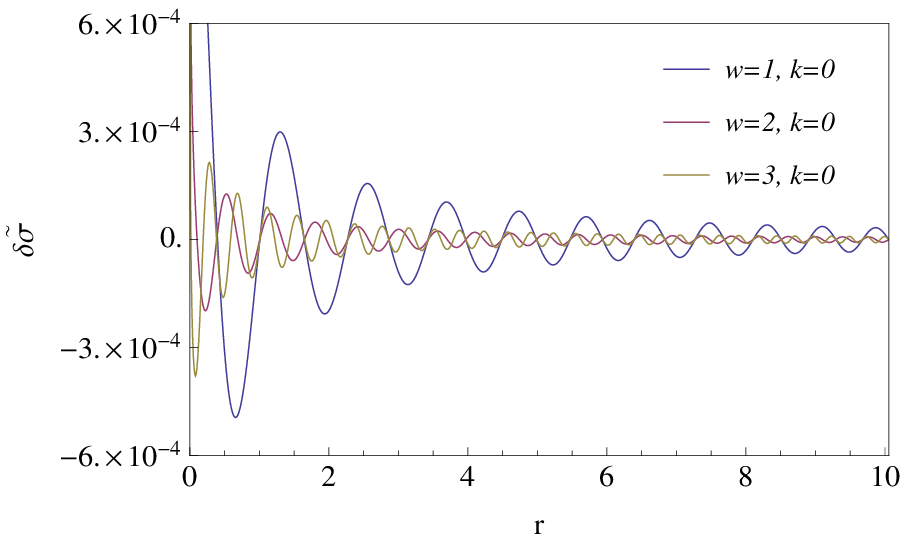}
\label{fig1}
\end{center}
\end{figure}
\vspace{-8mm}
\begin{small}
FIGURE 2. Perturbed energy density profiles $\delta \tilde{\sigma}=\sqrt{g_{zz}}\delta \sigma$ for different values of $w$ using Eq. \eqref{75} (upper panel) and Eq. \eqref{bar} (lower panel) in the isotropic Schwarzschild thin disk.
\end{small}

On the other hand, in Fig. 3 we present the profile of the perturbed pressure. When fixing $w$ and varying $k$, the amplitude for $\delta p$ has an oscillatory behavior similar to the one of $\delta \sigma$. From Fig. 3 it can be seen that the oscillation amplitude is not only smaller for Eq. \eqref{75} (upper panel) than for Eq. \eqref{bar} (upper panel), but also disappears in approximately the middle of the disk. It means that for the set of parameters here considered, the density and pressure in the disk are stable independently of the mass definition used.
\begin{figure}[H]
\begin{center}
\includegraphics[scale=0.85]{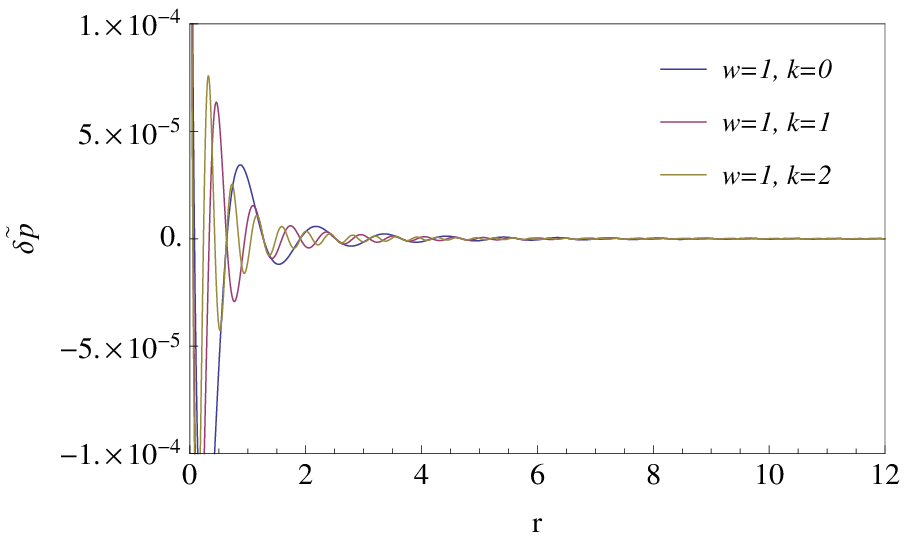}
\includegraphics[scale=0.85]{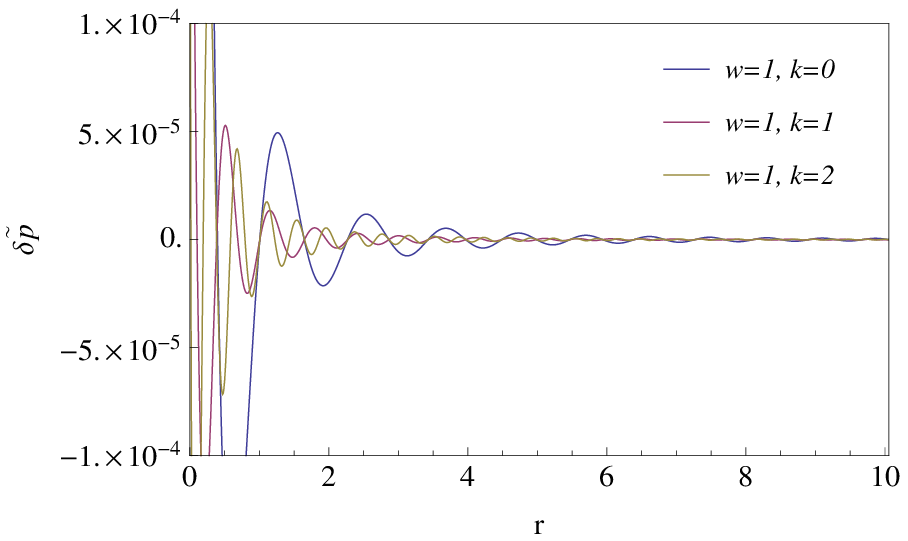}
\label{fig2}
\end{center}
\end{figure}
\vspace{-8mm}
\begin{small}
FIGURE 3. Perturbed pressure profiles $\delta \tilde{p}=\sqrt{g_{zz}}\delta p$ for the isotropic Schwarzschild thin disk using Eq. \eqref{75} (upper panel) and Eq. \eqref{bar} (lower panel). 
\end{small}

A full analysis of the system requires also a comparison between the amplitude of the perturbed velocities with the escape velocity of the constituents in the disk. In Fig. 4 we show the profiles for the perturbed radial $\delta U^{r}$  velocities using Eq. \eqref{75} (upper panel) and Eq. \eqref{bar} (lower panel), setting the respective cutoff radii for the 95.776\% of the total mass of infinite disk\footnote{We set this percentage because it gives place to unstabilities.}. Both velocities exhibit an increase in the frequency of oscillations with increasing $w$; moreover, we see that the envelopes of the oscillating functions increase at the external radial boundary, however, the perturbed radial velocity grows faster in the lower panel than in the upper panel. 

Concerning the escape velocity $v_{e}$, it is well known that it should be larger than the perturbed radial velocity, otherwise, the model will not have any astrophysical validity. The escape velocity required to overcome the force generated by a gravitational potential $\Phi$ can be calculated as $v_{e}=\sqrt{2\Phi}$, which from the weak-field approximation is given as $ 2\Phi=g_{t t}-1$. So, from \eqref{56}, we find 

\begin{equation}
\label{73}
v_{e}=\dfrac{2\sqrt{2mR}}{2R+m}.
\end{equation}

For the set of parameters $a=0.6$ and $m=0.5$, the resulting escape velocities for Eq. \eqref{75} and Eq. \eqref{bar} are $v_{e}=0.26$ and $v_{e}=0.28$, respectively. From Fig. 4 it is clear that using Eq. \eqref{75} the perturbed radial velocity is always less than the escape velocity (red dashed line), such that all particles remain inside the disk. Conversely, using Eq. \eqref{bar} the perturbed radial velocity can be greater than the escape velocity (red dashed line), meaning that the particles may escape of the disk. 

\begin{figure}[H]
\begin{center}
\includegraphics[scale=0.85]{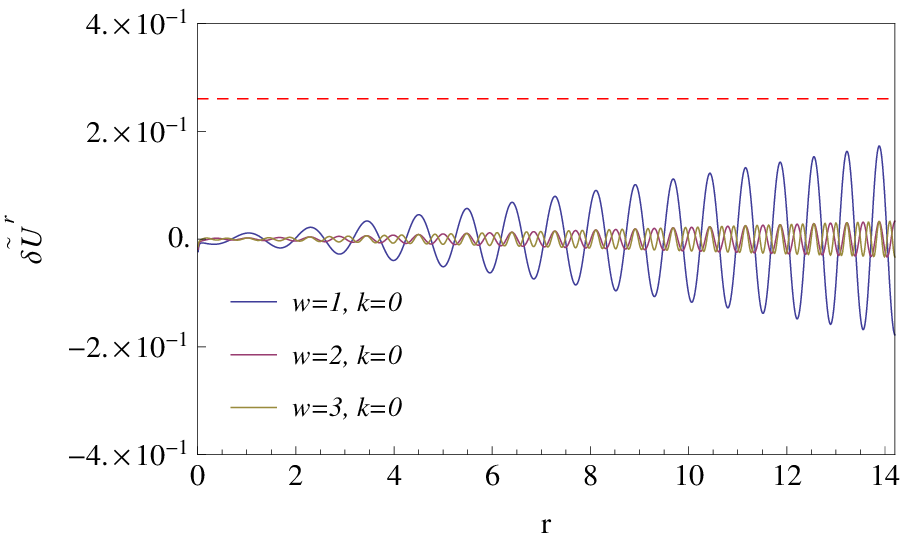}
\includegraphics[scale=0.85]{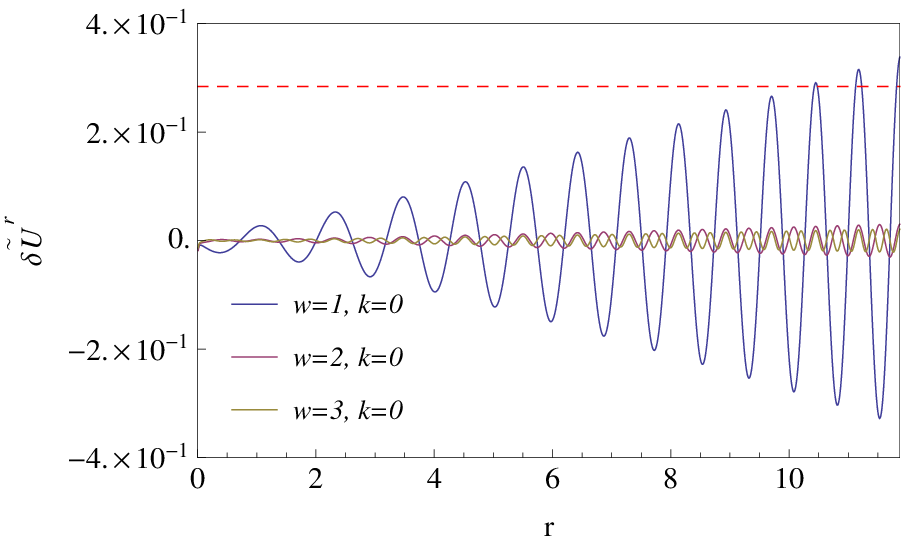}
\label{fig3}
\end{center}
\end{figure}
\vspace{-8mm}
\begin{small}
FIGURE 4. Profiles of the perturbed radial velocities $\delta\tilde{U}^{r}$ for different oscillation modes in Schwarzschild isotropic thin disk using Eq. \eqref{75} (upper panel) and Eq. \eqref{bar} (lower panel). The red dashed line denotes the value of the respective escape velocities. 
\end{small}

In Fig. 5 we show the profiles of the perturbed azimuthal velocities. In both cases the velocities are stable and some orders of magnitude smaller than the escape velocity. These results ratify that, under first order perturbations of the form \eqref{28} and using the Komar mass definition, the isotropic Schwarzschild thin disk has a stable behavior for a large set of parameters such that it can be used to describe astrophysical models.

\begin{figure}[H]
\begin{center}
\includegraphics[scale=0.85]{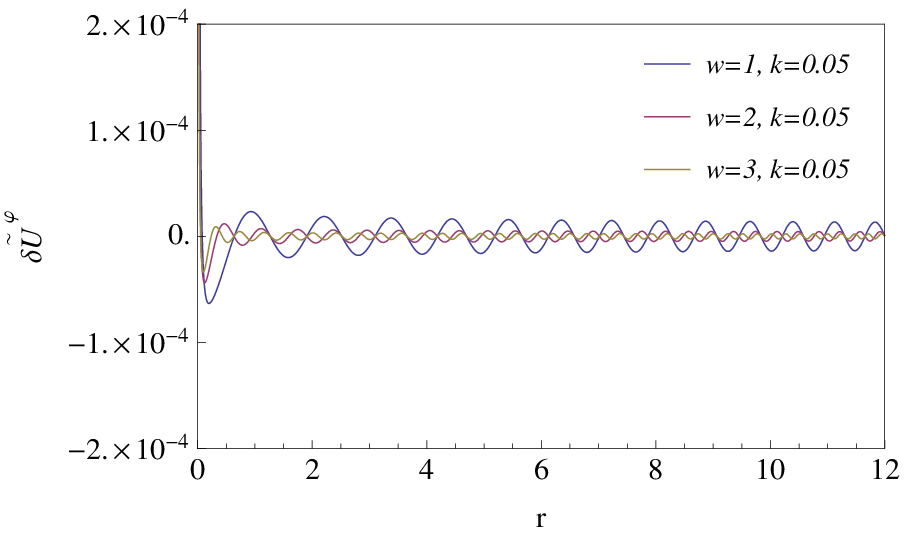}
\includegraphics[scale=0.85]{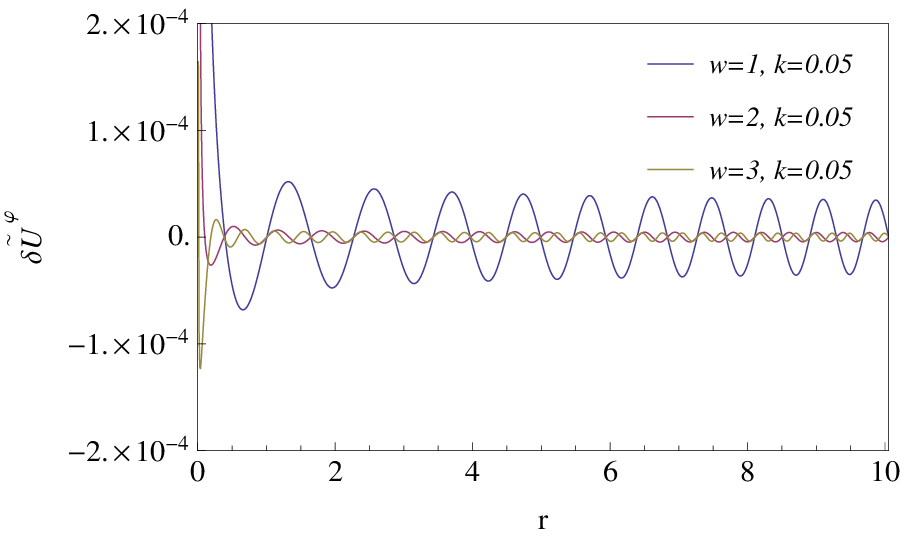}
\label{fig5}
\end{center}
\end{figure}
\vspace{-8mm}
\begin{small}
FIGURE 5. Profiles of the perturbed azimuthal velocities for different oscillation modes in Schwarzschild isotropic thin disk using Eq. \eqref{75} (upper panel) and Eq. \eqref{bar} (lower panel). 
\end{small}

\section*{Case 2: Chazy-Curzon Thin Disk}\label{fitting}

The Chazy-Curzon thin disk metric in Weyl coordinates is described as \citep{bicak1993-3}
\begin{equation}
\label{76}
ds^{2}=-e^{2\Phi}dt^{2} + e^{-2\Phi}r^{2}d\varphi^{2}+e^{2(\Lambda-\Phi)}(dr^{2}+dz^{2}),
\end{equation}
where the metric functions $\Phi$ y $\Lambda$ are given by
\begin{equation}
\label{78}
\Phi=-\dfrac{m}{R}, \quad \quad \quad \Lambda=-\dfrac{m^{2}r^{2}}{2R^{4}},
\end{equation}
with $R^{2}= r^{2}+\left(\lvert z \rvert + a\right)^{2}$. From \eqref{76} and \eqref{78}, and equations \eqref{51}, \eqref{52}, and \eqref{53}, we obtain the expressions for the pressure and the surface energy density of the Chazy-Curzon thin disk, 
\begin{equation}
\label{79}
\sigma= \dfrac{m a }{2\pi R^{3}_{0}}\left[1-\dfrac{m r^{2}}{R^{3}_{0}}\right]e^{2(\Phi_{0}-\Lambda_{0})},
\end{equation}
\begin{equation}
\label{80}
p_{\varphi}= \dfrac{m^{2} a}{2\pi R^{4}_{0}}\dfrac{r^{2}}{R^{2}_{0}}e^{2(\Phi_{0}-\Lambda_{0})}, \quad \quad p_{r}= 0,
\end{equation}
where $R_{0}$, $\Lambda_{0}$ and $\Phi_{0}$ are the functions evaluated at $z=0$. 

\begin{figure}[H]
\begin{center}
\includegraphics[scale=0.85]{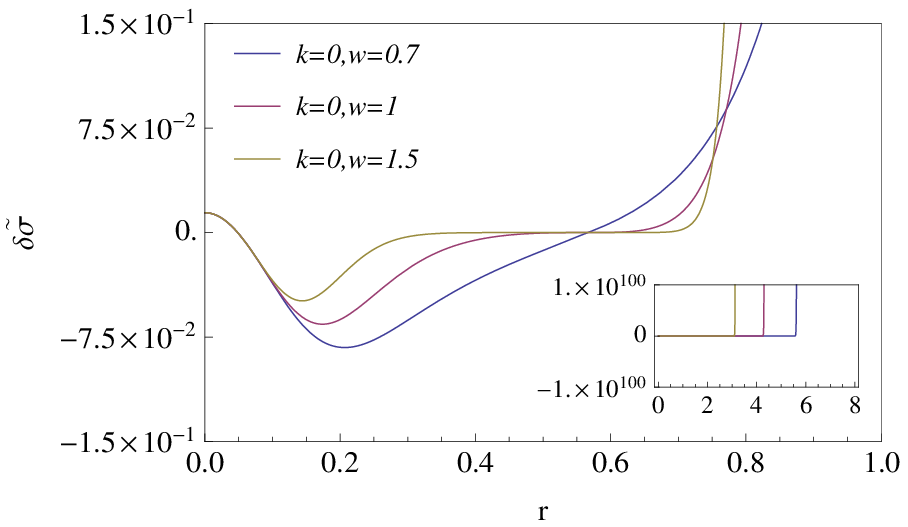}
\includegraphics[scale=0.85]{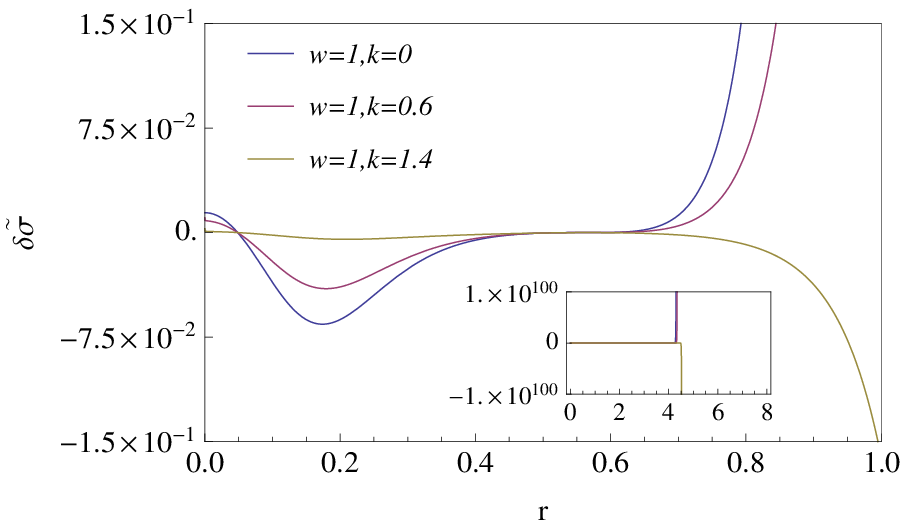}
\label{fig1}
\end{center}
\end{figure}
\vspace{-8mm}
\begin{small}
FIGURE 6.  Profiles of the perturbed energy density $\delta \tilde{\sigma}$ in the Chazy-Curzon thin disk using Eq. \eqref{75}, for different values of $w$ and $k$. The parameters have been set as $a=0.4$ and $m=0.5$. The insets show that for finite values of $r$, the perturbations tend to infinity.
\end{small}

Substituting equations \eqref{78}, \eqref{79} and \eqref{80}, into the Komar mass equation \eqref{75}, we obtain, accordingly, $M_{k} = m$. A particular set of values that satisfy the energy conditions are $a=0.4$ and $m=0.5$, in which case $r_c\approx 8$. In Fig. 6 we show the numerical solution to the differential equation  \eqref{68} with boundary conditions at $r=0$ and $r=r_{c}=8$. As can be noted, regardless of the values of $w$ (upper panel) and $k$ (lower panel), the profiles of the perturbed energy density grow rapidly to infinity. To rule out that the above effect is due to the chosen values of $a$, we perform an analysis varying $\delta \sigma$ in the interval $0\leq m/a\leq 1.3 $, observing the same tendency. The same procedure was performed using the total surface mass definition, obtaining the same behavior. These results show that the energy density, and hence the pressure and velocities, exhibit instabilities for this model and are independent of the mass definition used in the calculations. 

\section*{Case 3: Zipoy-Voorhees Thin Disk}\label{fitting}

The Zipoy-Voorhees thin disk metric in Weyl coordinates has the form  \citep{Ujevic2004}
\begin{equation}
\label{66}
ds^{2}=-e^{2\Phi}dt^{2} + e^{-2\Phi}r^{2}d\varphi^{2}+e^{2(\Lambda-\Phi)}(dr^{2}+dz^{2}),
\end{equation}
with metric functions $\Phi$ and $\Lambda$, given by
\begin{equation}
\label{88}
\begin{aligned}
& \Phi=\dfrac{m}{b-a}\ln\left[\dfrac{R_{a}+\lvert z \rvert+a}{R_{b}+\lvert z \rvert+b}\right],\\ \\
& \Lambda=\dfrac{2m^{2}}{(b-a)^{2}}\ln\left[\dfrac{(R_{a}+R_{b})^{2}-(b-a)^{2}}{4 R_{a}R_{b}}\right],
\end{aligned}
\end{equation}
where $R^{2}_{a}= r^{2}+\left(\lvert z \rvert + a\right)^{2}$, \  $R^{2}_{b}= r^{2}+\left(\lvert z \rvert + b\right)^{2}$ and $b\geqslant a$. Hence, by applying the same procedure used for the previous disks models, we obtain expressions for the surface energy density and the azimuthal pressure,
\begin{equation}
\label{90}
\begin{aligned}
\sigma=& - \dfrac{m^{2}e^{-2(\Lambda_{0}-\Phi_{0})}}{2\pi (b-a)^{2}} \left[\dfrac{2(R_{a0}+R_{b0})\left(\dfrac{a}{R_{a0}}+\dfrac{b}{R_{b0}}\right)}{(R_{a0}+R_{b0})^{2}-(b-a)^{2}} \right.\\
& \left. - \left(\dfrac{a}{R^{2}_{a0}}+\dfrac{b}{R^{2}_{b0}}\right)\right] +\dfrac{me^{-2(\Lambda_{0}-\Phi_{0})}}{2\pi (b-a)}\left(\dfrac{1}{R_{a0}}-\dfrac{1}{R_{b0}}\right),
\end{aligned}
\end{equation}
\begin{equation}
\label{91}
\begin{aligned}
p_{\varphi}= & \dfrac{m^{2} e^{-2(\Lambda_{0}-\Phi_{0})}}{2\pi (b-a)^{2}} \left[\dfrac{2(R_{a0}+R_{b0})\left(\dfrac{a}{R_{a0}}+\dfrac{b}{R_{b0}}\right)}{(R_{a0}+R_{b0})^{2}-(b-a)^{2}}- \right. \\
& \left. \left(\dfrac{a}{R^{2}_{a0}}+\dfrac{b}{R^{2}_{b0}}\right)\right],
\end{aligned}
\end{equation}

with $R_{a0}$, $R_{b0}$, $\Lambda_{0}$ and $\Phi_{0}$ the respective functions evaluated at $z=0$. Just like in the Chazy-Curzon model, the thin disk described by the Zipoy-Voorhees metric has no radial pressure.
Replacing \eqref{88}, \eqref{90} and \eqref{91} into equation \eqref{75}, we can calculate the Komar mass for the Zipoy-Voorhees thin disk as $M_{T} = m$.

To be consistent with the previous models, the calculations were performed for a particular set of values that satisfy the energy conditions, $m=0.5$,  $a=1$ and $b=2.15$, such that the cutoff radius takes the value $r_c\approx 31$. In the same way as cases 1 and 2, in Fig. 7 we show the numerical solution to the differential equation  \eqref{68} with boundary conditions at $r=0$ and $r=rc =31$. 

It can be seen from Fig. 7 that there exist instabilities for the Zipoy-Voorhees thin disk strongly amplified before they reach 10\% of its cutoff radius, and regardless of the angular frecuency $w$ (upper panel) or wave number $k$ (lower panel) the perturbed energy density tends to infinity. The results for the Zipoy-Voorhees thin disk when using the total surface mass are not shown because they exhibit practically the same behavior than using the Komar mass. 
The independence of the results with the particular chosen values of $a$ and $b$, was also analyzed observing the same tendency. Given the instability in the surface energy density, the other thermodynamic variables of the thin disk also are unstable.

\begin{figure}[H]
\begin{center}
\includegraphics[scale=0.85]{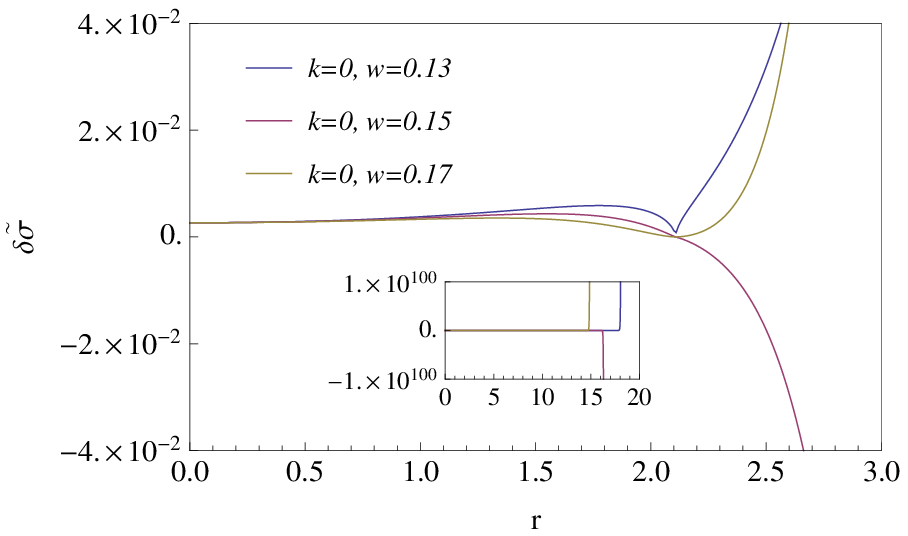}
\includegraphics[scale=0.85]{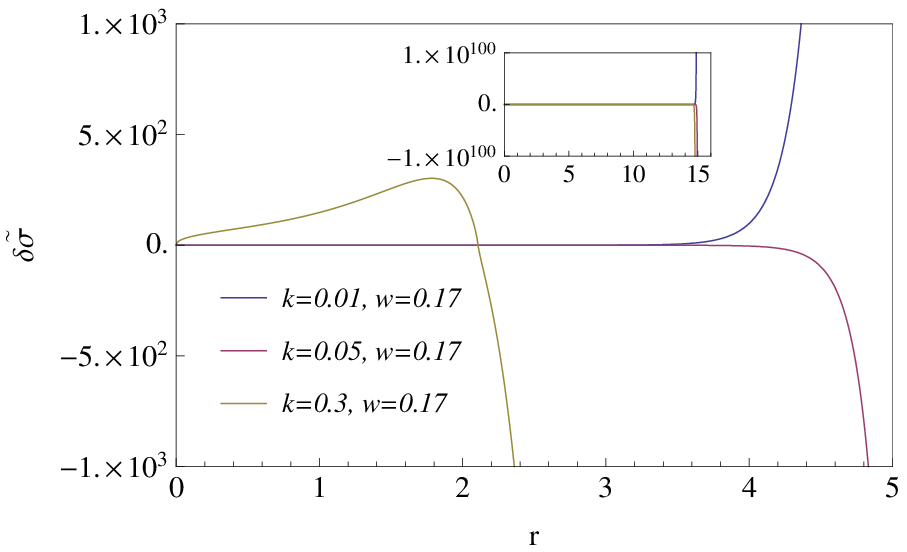}
\label{fig1}
\end{center}
\end{figure}
\vspace{-8mm}
\begin{small}
FIGURE 7.  Profiles of the perturbed  energy density  $\delta \tilde{\sigma}$ for different values of $w$ (upper panel) and $k$ (lower panel) in the Zipoy-Voorhees thin disk. The parameters have been set as $a=1$, $b=2.15$ and $m=0.5$. The inset shows the tendency of the curves for larger scales.
\end{small}

\section*{Concluding Remarks}\label{discuss}
 In this paper, we have reviewed the results obtained by \cite{Ujevic2004} on the stability of three particular models, using an alternative mass definition, {\it i.e.}, the Komar mass. This formal concept of mass can be defined in any stationary spacetime, so it is applicable to all three particular thin disk models under consideration: the isotropic Schwarzschild, Chazy-Curzon, and Zipoy-Voorhees metrics. Assuming that the disks are made of a perfect fluid, we found the differential equation for the perturbed energy density and the non- zero components of the surface energy-momentum tensor. The differential equation was numerically solved, defining a cutoff radius $r_{c}$, such that the matter up to $r_{c}$ is approximately 95\% of the mass of the infinite disk and the remaining 5\% of matter is distributed from $r_{c}$ to infinity. Moreover, we assumed that at the center of the disk the perturbation equals 10\% of the unperturbed energy density, and at the external cutoff radius the perturbation vanishes.

Once we derived the expressions for the pressure components, the surface energy density, and the metric functions for each particular case, we calculated the Komar mass. As a general result, we find that the mass parameter in each one of the metrics equals the Komar mass for the disk, while by using the total surface mass definition, as is the case of the reviewed paper, the total mass depends on the parameters of the  DCR method. The use of the Komar definition lets us set the physical parameters that best fit the expected energy density profile in a realistic model and simultaneously satisfy the energy conditions. With this result, the cutoff radius only depends on the thermodynamic variables and the free parameters of each metric. 

As the main finding, we found that the cutoff radius is larger for the Komar mass definition than for the total surface mass. This result let us to increase the number of parameters that give place to stable Schwarzschild thin disk models. On the other hand, the Chazy-Curzon and Zipoy-Voorhees thin disk models are not stable under first- order perturbations, because the thermodynamic variables and fluid velocities tend to infinity for finite values of the radial coordinate. Such instabilities are a consequence of the lack of radial pressure and are not related to the definition of mass. We also have shown that the infinite radial extension of the disk can be the reason for the instability, as hypothesized by Ujevic {\it et al.}, nevertheless, some instabilities can be artificially introduced into the model due to the use of a non-appropriate mass definition in the calculations.

{\bf Acknowledgments.} We would like to thank the anonymous referee for useful comments and remarks, which improved the presentation of the paper. This research was partially supported by VIE-UIS under grant numbers 1822 and 1785, and COLCIENCIAS under grant number 8840. EAB would like to thank COLCIENCIAS for their support through the program: Jóvenes Investigadores e Innovadores 2014. FLD acknowledges financial support from the Universidad de los Llanos provided under grant Commission: Postdoctoral Fellowship Scheme.

{\bf Conflict of interest.} The authors declare that they have no conflict of interest.

\bibliographystyle{chicago}

\renewcommand{\refname}

\end{multicols}
\end{small}
\end{document}